\documentclass[12pt]{article}
\usepackage[utf8]{inputenc}
\usepackage{graphicx}
\usepackage[utf8]{inputenc}
\usepackage{amsmath}
\usepackage{amsfonts}
\usepackage{amssymb}
\usepackage{graphicx}
\usepackage{setspace}
\usepackage{authblk}
\usepackage{caption}
\usepackage{float}
\usepackage{cite}
\usepackage[left=2cm,right=2cm,top=2cm,bottom=2cm]{geometry}
\usepackage{color,soul}
\usepackage[dvipsnames]{xcolor}
\title{\textbf{Coupling between DBI dark energy model and $f(Q)$ gravity and its effect on Condensed Body Mass accretion}}
\author[1]{Alokananda Kar}
\author[2]{Shouvik Sadhukhan}
\author[3]{Ujjal Debnath}
\affil[1]{\small{Department of Physics; Indian Institute of Technology (ISM), Dhanbad; Police Line Road, Main Campus IIT (ISM, near Rani Bandh, Hirapur, Sardar Patel Nagar, Dhanbad, Jharkhand 826004, India \par
Corresponding Author Email: alokanandakar@gmail.com}} \affil[2]{\small{Department of Physics; Indian Institute of Space Science and Technology, Thiruvananthapuram; Valiamala Road, Valiamala, Kerala 695547 \par Email: shouvikphysics1996@gmail.com}}
\affil[3]{\small{Department of Mathematics, Indian Institute of Engineering Science and Technology, Shibpur, Howrah-711 103, West Bengal, India.\par Email: ujjaldebnath@gmail.com}}
\date{}
\begin{document}
\maketitle

\begin{abstract}
We have studied the reconstruction formalism of
the Dirac-Born-Infeld (DBI)-essence scalar field model in the
background of non-metric gravity or $f(Q)$ gravity which is
controlled by the deformation or non-metricity scale $Q$. The
deformation is caused by the fifth dimension of wrapped compact
spaces in brane cosmology. The fifth dimension of that wrapped
space is controlled by the D-brane tension $T(\phi)$. Hence, we
have reconstructed the formalism of DBI scalar field energy
densities and pressures using the coupling of deformation geometry
and brane scalar fields, i.e., $f(Q)$ and DBI-Scalar fields. The
accretion of dark energy onto black holes and wormholes and its
critical analysis have been studied with this reconstructed model.
The validity of energy conditions has been studied. We have
assumed four types of singularity resolutions of the scale factor
to investigate the nature of the black hole and wormhole masses
due to accretion. Graphically, the physical quantities like mass,
kinetic and potential energies have been analyzed for DBI-essence
model in the background of $f(Q)$ gravity.\\


\end{abstract}

\section{Introduction}
Over the years, several problems have come up while discussing the
evolution of the universe, and several new theories have been put
forward to explain them. The theory of inflation was introduced to
explain the horizon problem, flatness problem, and magnetic
monopole problem. After the discovery of the accelerated expansion
of the universe, several theories were introduced to explain this
accelerated expansion since we know, of the attractive nature of
gravity, accelerated expansion of gravity is not a possibility, as
predicted by the Raychaudhuri equations. To explain this repulsive
nature, we need the idea of negative pressure. Since no known
matter can have negative pressure, the idea of dark energy (DE)
was introduced, assuming that this form of energy can give the
required negative pressure, which is necessary for accelerated
expansion. A positive cosmological constant is the oldest form of
dark energy. But this model had many shortcomings. When the
experimental and theoretical data for vacuum energy differed by
orders of $10^{120}$, scientists came up with the idea of
dynamical dark energy models. Till now, several models of Dark
energy have been proposed, such as the Dark fluid model (Chaplygin
gas model, Tachyon model), scalar field model (DBI-essence model,
Quintessence model) and holographic model. These dark energy
models are efficient in providing a proper definition of negative
pressure \cite{1,2,3,4,5}.\par Modified gravity is a type of
alternative gravity theory that takes higher-order terms in
Einstein's action.

The theories of modified gravity are used to explain accelerated
expansion. Several types of modified gravity are available in the
literature. All these candidates of modified gravity have been
brought depending upon the scheme of higher-order terms taken in
Einstein's action. $f(R)$ gravity, Gauss-Bonnet gravity,
Teleparallel gravity and gravity with non-metricity are all
included in the list of modifications of
gravity.\cite{11,12,13,14,15,16,17,18,19,20,21}\par
The modified gravity models discuss the
complexities of the higher ordered Ricci tensor and scalar based
action with their effect on cosmic dynamics \cite{34}. Recent
literature discusses the decompositions of these complexities
using tensor decomposition calculus mechanism \cite{1}.
Teleparallel modifications are some of them \cite{2}. In such a
mechanism, the complex or higher ordered Christoffel symbol or
affine connection can be decomposed to Christoffel connection for
Einstein action, contortion, and the deformation of frame
\cite{35,36,37}. This decomposition formalism makes the geometry
non-metric, and thus, we can generate the non-metric gravity or
$f(Q)$ gravity \cite{34,35}.
Here, the term $Q$ represents the non-metric scalar which can be
written as
$Q=-\frac{1}{4}Q_{\alpha\beta\mu}Q^{\alpha\beta\mu}+\frac{1}{2}Q_{\alpha\beta\mu}Q^{\beta\mu\alpha}+\frac{1}{4}Q_{\alpha}Q^{\alpha}-\frac{1}{2}Q_{\alpha}Q^{\alpha}$
\cite{34,35}. The non-metric tensor can be written as
$Q_{\mu\nu}=\nabla_{\alpha}g_{\mu\nu}$. The non-metricity is
basically introduced when the Christoffel connection decomposed,
and the deformation tensor is generated as the form of
$L^{\alpha}_{\mu\nu}=Q^{\alpha}_{\mu\nu}-Q_{(\mu,\nu)}^{\alpha}$
\cite{34,35}. Hence, the extra deformation in geometry due to the
higher order modifications of Ricci tensors and metric tensors
provides the non-metric counterpart of modified geometric physics
\cite{3}. When this deformation causes the negative pressure of
universe expansion, this formalism can be used as an alternative
to exotic energy-based cosmology or dark energy \cite{4}. The
transformation from non-decomposed geometry to decomposed geometry
and the decay of deformation, as well as contortion, can provide
the early time accelerated expansion or inflation of the universe
\cite{5}.\par Among the list of dark energy
scalar field models, the DBI-essence model is the one that
contains the information regarding the stringy cosmological
geometry \cite{6}. According to this model, the cosmic evolution
is controlled by the D3-Brane in the warped throat region of the
compact bulk space \cite{7}. It basically modifies the invariant
properties of the geometric line element, and that is why the
modulation factor of Einstein Hilbert action, i.e. the $\sqrt{-g}$
gets modifications due to the fifth-dimensional warp-factor like
$\sqrt{-det(g_{\mu\nu})}\rightarrow\sqrt{-det(A_{\mu\nu})}$
\cite{8}. The wrap factor basically modifies the metric tensor
$g_{\mu\nu}$ \cite{9}. This modification of metric tensor comes
due to  D3-Brane tension and its potential $T(\phi)$ \cite{10}.
Using that warp factor, the modulation factor of action can be
expressed as
$\sqrt{-det(A_{\mu\nu})}=\sqrt{-g}T(\phi)\sqrt{1-\frac{\dot{\phi}^2}{T(\phi)}}$
\cite{6,7}. The tension potential modifies the scalar field
potential of p-Brane also due to the opposing nature of $T(\phi)$
\cite{7,8}. Hence, the modified potential can be written as
$V_{modified}(\phi)=V(\phi)-T(\phi)$ \cite{8,9}. The kinetic
energy is used to modify the modulation factor of action, and
metric tensor \cite{9,10}. The kinetic energy of this scalar field
is called non-canonical kinetic energy \cite{6,7,8}. It doesn't
support the canonical formalism of scalar field phase space
dynamics in usual manners \cite{7,8,9}. Hence, the final
Lagrangian for DBI-Essence scalar fields can be written as
$\sqrt{-g}(T(\phi)\sqrt{1-\frac{\dot\phi^2}{T(\phi)}}+V(\phi)-T(\phi))$
\cite{8,9,10}. Thus, the transformation of such non-canonical
kinetic energy into canonical kinetic energy through D3-Brane
decay can provide a transformation from warp space-time to usual
Einstein space-time. This can be expressed using metric tensor
transformation as
$\sqrt{-det(A_{\mu\nu})}\rightarrow\sqrt{-det(g_{\mu\nu})}$. This
phenomenon can provide inflationary expansion or early time
accelerated expansion of the universe. Thus, the decay of brane
tension $T(\phi)$ provides the alternative scalar field-based dark
energy models or exotic type matter field-dominated universe.
\cite{6,7,8,9,10}\par The deformation we
discussed in the decomposed Christoffel Connection can be caused
by the D3-Brane of warped compact bulk space-time structure. The
fifth dimension type presence of D3-Brane in between the p-branes
can cause the scalar field cosmic matter tension. The tension on
the usual scalar field due to this fifth dimension on the warped
system can provide the geometric deformation with non-metricity.
Hence, both the decay of brane tension $T(\phi)$ and the
deformation factor of Christoffel connection can transfer the
universe from the primordial phase to the late time phases. This
transformation causes inflation. Hence, the DBI-essence model
provides the dark universe through the cosmic matter scalar field
point of view, whereas the $f(Q)$ gravity provides the same
through the geometric point of view. Thus, we can couple them to
introduce a cause behind the deformation of geometry found from
the decomposition of the Christoffel Connection. The coupled
Lagrangian hence can be considered as $L_Q+L_{\phi}$. The
modification of the scalar field model using the D3-brane fifth
dimension is basically the modifications of matter lagrangian
through scalar field energy point of view. Hence, the effective
system should be coupled as $L_Q+L_{\phi}+L_m$.\par
The mass accretion of blackholes and wormholes
mainly increases the mass of the blackholes accretion disc. In
general, the compact bodies form the blackholes when their mass
crosses a certain critical value beyond which it fails to exert
enough hydrodynamical forces to repel the gravities due to masses.
The gravitational collapse provides a boundary which splits the
time outsides with the space like an inside zone. This splitting
region is called the horizon. The horizons are also called null
surfaces where all the killing vectors get null magnitudes, and
hence, the time seems infinite w.r.t any outside observer for
penetration by any arbitrary objects. Thus, the increase in mass
of the black hole producing a compact body using mass accretion is
impossible in physics. Hence, the accretion disc geometry evolves
with such mass accretion. The accretion disc provides drag forces
due to its fluid dynamics and wrapped geometry on the incoming
masses. The evolution of $f(Q)$ gravity and wrapping tension or
D-brane tension evolution provides a modification on the mass
accretion and its rate in the accretion disc. Hence, the drag
force, fluid dynamics and the wrapping geometry in the accretion
disc are indirectly modified by the evolution of modified gravity
and DBI-essence scalar field dynamics. Thus, mass accretion
becomes important for present universe black hole models. The
wormhole geometry can be evolved with the geometry of the outside
universe. The wormholes are the passage of masses through some
time-like zone bounded by two throats with space-like boundaries.
Hence, the evolution of the geometry of the universe and scalar
fields, as well as the mass accretion, provides a modification to
the throats and passage geometry. Thus, the mass accretion is
linked with the geometry of blackholes and
wormholes.\cite{21,22,23,24,25}\par The
primordial black holes are caused by the topological factors of
the early-time inhomogeneous and anisotropic geometry. Hence,
primordial black holes are called topological black holes. The
topological black holes also provide horizon and deep curvatures
but are caused by the early universe phase deformations. Hence,
the evolution of such deformations can also provide the evolution
of the structures and equivalent masses of those topological black
holes. The mass accretion in those phases is the evolution of
geometries, and hence, this can modify the primordial black hole
systems.\cite{22}\par The energy conditions for
the universe must have to be specified due to the expansion
scheme. The NEC must be satisfied, and SEC must be violated.
Hence, the evolutions of the energy conditions provide the idea of
the expansion scheme of the universe, which modifies the nature of
black holes, as told above. Thus, energy conditions are linked
with the blackholes and wormholes indirectly.\cite{25}\par
In our work, we have used the local static
solutions of blackholes and wormholes from Einstein's universe. We
have used Schwarzschild solutions of compact bodies as blackholes
and Morris-Thorne wormhole metrics in our work. The fundamental
difference between the compact bodies found in modified gravity
with our work is that we have used the compact bodies of
Einstein's universe and modified them with mass accretion using
higher order deformations. The modified gravity modifies the local
solutions of blackholes and wormholes where the line elements are
different from local solutions in the Einstein universe.
\cite{11,12,13,14,15,16,17,18,19,20,21}\par The
fundamental difference in mass accretion between the primordial
blackholes and present-time blackholes is their nature the origin
of them. The origin of present-time blackholes is the
gravitational collapse, whereas the origin of primordial black
holes is the topological deformations of space-time geometry due
to inhomogeneous fluid and based cosmic evolutions. Hence, the
mass accretion and its effects on those two blackholes are
different.\cite{11,12,13,14,15,16}\par In this paper, we have
concentrated on DBI-essence model of dark energy. This model takes
Brane Bulk tension into account, which was not included in the
previous scalar field models. Hence this is a more generalised
model. Martin and Yamaguchi [6], in their work, have considered
the DBI-essence field as the dark energy scalar field, where the
kinetic energy term has the DBi form. Because of this, the kinetic
term is non-canonical. The potential arises from the internal
tension between D-branes, which controls the geometry of the
warped region in compact space. Therefore, the DBI-essence model
is a bit different from other slow-roll inflation models.
\cite{6,7,8,9,10}\par In this work we have focused on $f(Q)$
gravity which is a teleparallel modification of symmetric gravity.
The metric-affine variational principle is used to obtain the
symmetric teleparallel theory of gravity. The motivation for
working in $f(Q)$ gravity is that the equations of motions are in
second-order, which are easier to solve. Recently in some
observational analysis \cite{7} it was found that the well-known
$f(Q)$ gravity may challenge Einstein's gravity with its
viability. They have shown their analysis with Supernovae type Ia
(SNIa), Baryonic Acoustic Oscillations (BAO), cosmic chronometers
(CC), and Redshift Space Distortion (RSD) data that motivated us
to further study on $f(Q)$ gravity.
\cite{11,12,13,14,15,16,17,18,19,20,21}.\par Here, we have
investigated the coupling between the DBI-Scalar field and $f(Q)$
gravity. We will investigate the changes in the mass accretion
rate of condensed objects caused due to these modifications. Using
the Holographic technique, Babichev et al. \cite{31} calculated
the accretion of phantom DE into a Schwarzschild black hole and
observed that mass decreases to zero near the big rip singularity.
In this paper, we have reconstructed the DBI-essence model with
modified gravity and investigated its effect on mass accretion. We
have also provided the analysis of thermodynamics energy
conditions to bring the acceleration in our model \cite{22, 23,
24}. \par The paper is organized as follows: In sections 2 and 3,
we discuss the basics of $f(Q)$ gravity and the DBI-essence model,
respectively. In section 4, we have introduced coupling between
the modified gravity and dark energy model for the reconstruction
mechanism. In section 5, we discuss the thermodynamic energy
conditions. Section 6 contains the basic discussion on mass
accretion rate variation \cite{26,27}. In section 7, we take four
types of scale factors \cite{24, 25,26,27,28,29,30,31,32} and have
investigated all the above-mentioned conditions. The physical
analysis of the models has been discussed in section 8. Finally,
we draw some conclusions from the work in section 9.

\section{Overview of $f(Q)$ Gravity model}

The action of the universe governed by $f(Q)$ is given by \cite{1,2,3,4,5}
\begin{equation}
    S =\int{\frac{1}{2}f(Q)\sqrt{-g}d^4x}+\int{L_m\sqrt{-g}d^4x}
\end{equation}
where $f(Q)$ is arbitrary function of non-metricity $Q$ and
$L_m$ is the matter Lagrangian.\\
The non-metric tensor is defined by
\begin{equation}
    Q_{\lambda\mu\nu}= \triangledown_\lambda g_{\mu\nu}
\end{equation}
We'll use FRW curvature free line element as
\begin{equation}\label{3}
    ds^2=dt^2-a^2(t)(dx^2+dy^2+dz^2)
\end{equation}
The energy-momentum tensor is $T_{\mu\nu}=(\rho_m+p_m )u_\mu
u_\nu-p_m g_{\mu\nu}$ where $\rho_{m}$ and $p_{m}$ are
respectively the energy density and pressure of matter. The effective energy momentum tensor can be written as $(T_{\mu\nu})_{eff}=(\rho_{eff}+p_{eff} )u_\mu
u_\nu-p_{eff} g_{\mu\nu}$. Here we
have chosen $8\pi G =c= 1$. The trace of non-metric tensor with
respect to line element given by (\ref{3}) is
\begin{equation}
    Q=6H^2
\end{equation}
where $H=\dot{a}/a$ is the Hubble parameter. The Friedmann
equations describing the universe are
\begin{equation}\label{5}
    3H^2=\frac{1}{2f_Q}(\rho_m-\frac{f}{2})
\end{equation}
and
\begin{equation}\label{6}
    2\dot{H}+3H^2= \frac{1}{2f_Q}(-p_m-\frac{f}{2}-2\dot{f_Q}H)
\end{equation}
where $\rho=\rho_{m}+$ $f(Q)$ gravity energy density and $p=p_{m}+$ $f(Q)$ gravity pressure.

\section{Overview of DBI-essence Dark Energy}
DBI-essence model starts from the following action \cite{9}
\begin{equation}
    S=\int{d^{4}x\sqrt{-g}(T(\phi)\sqrt{1-\frac{\dot\phi^2}{T(\phi)}}+V(\phi)-T(\phi))}
\end{equation}
The dynamical nature of scalar field can be reproduced with the
Klein-Gordon equation as follows:
\begin{equation}
    \ddot{\phi}-\frac{3T'(\phi)}{2T(\phi)}\dot{\phi}^2+T'(\phi)
    +\frac{3}{\gamma^2}\frac{\dot{a}}{a}\dot{\phi}+\frac{1}{\gamma^3}(V'(\phi)-T'(\phi))=0
\end{equation}
where, $\gamma=\sqrt{1-\frac{\dot{\phi}^2}{T(\phi)}}$ provided
$\dot{\phi}^{2}<T(\phi)$. For simplicity, let us assume
$V(\phi)=T(\phi)=n\dot{\phi}^2$ with $n>1$. So
$\gamma=\sqrt{\frac{n}{n-1}}$. \par From the modified Klein-Gordon
equation, we get the kinetic energy of scalar field as follows:
\begin{equation}
    \frac{1}{2}\dot{\phi}^2=\frac{1}{2}\sqrt{\frac{n-1}{n}}[\rho_{\phi}+p_{\phi}]
\end{equation}
The scalar field potential should have the form as follows:
\begin{equation}
     V(\phi)=T(\phi)=\sqrt{n(n-1)}[\rho_{\phi}+p_{\phi}]
\end{equation}
The scalar field component of the coupled action, we get the
pressure and energy density for DBI-essence model as follows:
\begin{equation}
    \rho_{\phi}=(\gamma-1)T(\phi)+V(\phi)
\end{equation}
And,
\begin{equation}
    p_{\phi}=(1-\frac{1}{\gamma})T(\phi)-V(\phi)
\end{equation}

These scale factor dependent functions of density and pressure can
provide us with information about DBI-essence dark energy density
and pressure.

\section{Coupling between $f(Q)$ and DBI-essence model}

The reconstruction mechanism is done by introducing coupling
between the reconstructing system $f(Q)$ and the reconstructed
system (DBI-essence). In our calculation, the DBI-essence scalar
field is to be reconstructed. So, the coupled action can be
written as follows:
\begin{equation}
    S=\int{d^4 x\sqrt{-g}[\frac{f(Q)}{2}+T(\phi)\sqrt{1-\frac{\dot{\phi}^2}{T(\phi)}}+V(\phi)-T(\phi)+L_m]}
\end{equation}
So the Friedmann equations can be written as follows:
\begin{equation}\label{16}
    3H^2=\frac{1}{2f_Q}(\rho_m+\rho_{\phi}+\rho_Q)
\end{equation}
and
\begin{equation}\label{17}
    3H^2+2\dot{H}=\frac{1}{2f_Q}(-p_m-p_{\phi}-p_Q)
\end{equation}
From equation (\ref{5}) and (\ref{6}), we can write the
contributions of density and pressure of $f(Q)$ gravity for the
universe as
\begin{equation}\label{18}
    \rho_Q=-\frac{f}{2}
\end{equation}
and
\begin{equation}\label{19}
    p_Q=-f_{QQ}H\dot{Q}-\frac{f}{2}
\end{equation}
From the scalar field components, we know that the energy density
and energy density are as equations (\ref{16}) and (\ref{17}).
Using equations (\ref{18}) and (\ref{19}) in (\ref{16}) and
(\ref{17}) we get
\begin{equation}
    \rho_{\phi}=3H^2f'(Q)-\rho_m+\frac{f}{2}
\end{equation}
and
\begin{equation}
    p_{\phi}=-(3H^2+2\dot{H})f'(Q)+f_{QQ}H\dot{Q}-\frac{f}{2}
\end{equation}
where we have assumed cold dark matter with negligible pressure
($p_m \approx 0$). Detailed calculation and analysis of mass
accretion and other parameters will be done in the following
sections.

\section{Overview of thermodynamic energy conditions}

Raychaudhuri equations in cosmic fluid dynamics provides the
following energy conditions against the cosmic evolution.\cite{33}
\begin{center}
$\frac{\mathrm{d} \theta}{\mathrm{d} \tau}=-\frac{1}{3}\theta^2-\sigma_{\mu\nu}\sigma^{\mu\nu}+w_{\mu\nu}w^{\mu\nu}-R_{\mu\nu}u^{\mu}u^{\nu}$
\end{center}
And,\par
\begin{center}
$\frac{\mathrm{d} \theta}{\mathrm{d} \lambda}=-\frac{1}{3}\theta^2-\sigma_{\mu\nu}\sigma^{\mu\nu}+w_{\mu\nu}w^{\mu\nu}-R_{\mu\nu}n^{\mu}n^{\nu}$
\end{center}\par
where $\theta$ is the expansion factor, $n^{\mu}$ is the null
vector, and $\sigma_{\mu\nu}\sigma^{\mu\nu}$ and
$w_{\mu\nu}w^{\mu\nu}$ are, respectively, the shear and rotation
associated to the vector field $u^{\mu}$. Here $\lambda$ is affine
parameter. For attractive gravity we'll have the followings:
\begin{center}
$R_{\mu\nu}u^{\mu}u^{\nu}\geq0$ and $R_{\mu\nu}n^{\mu}n^{\nu}\geq0$
\end{center}\par
To set some nomenclature the energy conditions of general
relativity to be considered here are
\begin{itemize}
    \item [1] Null energy condition (NEC) or $\rho+p\geq0$
    \item [2] Weak energy condition (WEC) or $\rho\geq0$ and $\rho+p\geq0$
    \item [3] Strong energy condition (SEC) or $\rho+3p\geq0$ and $\rho+p\geq0$
    \item [4] Dominant energy condition (DEC) or $\rho\geq0$ and $-\rho\leq p\leq\rho$
\end{itemize}

\section{Overview of Black Holes and Wormholes Mass accretion}

In the case of mass accretion, the rate doesn't depend on the
metric or geometry of the black holes or wormholes. Although the
mass accretion of wormholes and black holes are different. We are
just mentioning the basic formulae to discuss the mass accretion
\cite{26,27,28,29,30}.\par The mass accretion of black holes can
be written as follows:
\begin{equation}\label{20}
    \dot{M}=-4\pi r^2 T_{0}^1=-4\pi A M^2 (\rho_{\phi}+\rho_m+p_{\phi}+p_m)
\end{equation}
The mass accretion of wormholes is given by
\begin{equation}\label{21}
    \dot{M}=4\pi r^2 T_{0}^1=4\pi B M^2 (\rho_{\phi}+\rho_m+p_{\phi}+p_m)
\end{equation}

We analyze the mass accretion due to the reconstructed scalar
field. For further calculation, we consider cold dark matter with
$p_m=0$.

\section{Different scale factor and Mass accretion}

In this section, we will introduce four different scale factors as
used in \cite{24} that can produce four different results of
scalar field energy densities, pressures, Scalar field kinetic
energies, and scalar field potential energies. The following
subsections will show all those parameters depending upon
different scale factors.\par Here we have used the functional form
of modified gravity as follows,
\begin{equation}
    f(Q)=\lambda Q^2
\end{equation}
where $\lambda$ is a constant.

\subsection{Scale factor $a(t)=a_0(a_1+nt)^m$}

Using the scale factor $a(t)=a_0(a_1+nt)^m$ in equations
(\ref{20}) and (\ref{21}) we get,
\begin{equation}
    \rho_{\phi}=-\frac{\rho_{m0}}{a^{3}_0 (a_1+nt)^3}+\frac{90m^4n^4\lambda}{(a_1+nt)^4}
\end{equation}

\begin{equation}
    p_{\phi}=-\frac{48m^3 n^4\lambda}{(a_1+nt)^4}+\frac{18m^4n^4\lambda}{(a_1+nt)^4}+\frac{12m^2 n^2 \lambda (\frac{4 m n^2\lambda}{(a_1+nt)^2}-\frac{6 m^2 n^2\lambda}{(a_1+nt)^2})}{(a_1+nt)^2}
\end{equation}

Here this scale factor is assumed from the idea of power-law type
functions. In general, this kind of scale factor dominates the
late time-expanding universe. Its range of time should be $t>0$ to
$t\rightarrow\infty$. All the constants $a_{0},~a_{1},~m,~n$ are
positive in nature.

So, those variables can be represented graphically as follows in
Figs. 1 - 4. From Fig. 1, we observe that the energy density
$\rho_{\phi}$ increases as time increases for $t<0$ (i.e., before
bounce) while $\rho_{\phi}$ decreases as time increases for $t>0$
(i.e., after the bounce) and near $t=0$, the value of density is
very high. On the other hand, before the bounce and after the
bounce, Fig.2 shows that pressure $p_{\phi}$ is very high negative
near $t=0$. From Figs. 3 and 4, we see that the kinetic energy and
potential energy decrease from a high positive value to a low
positive value as time goes on ($t>0$).

\begin{figure}[H]
\centering
\begin{minipage}[b]{0.4\textwidth}
    \includegraphics[width=\textwidth]{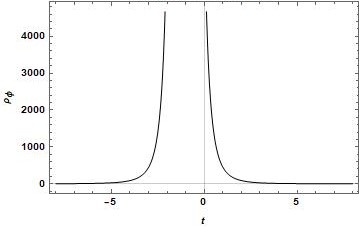}
    \caption{Plot of $\rho_{\phi}$ vs $t$ with $a_0 = 1$, $a_1 = 1$, $n = 1$, $\rho_{m0} = 1$, $m = 2$ and $\lambda = 5$}
\end{minipage}
\hfill
\begin{minipage}[b]{0.4\textwidth}
    \includegraphics[width=\textwidth]{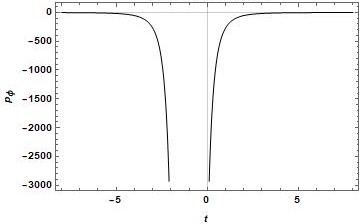}
    \caption{Plot of $p_{\phi}$ vs $t$ with $a_0 = 1$, $a_1 = 1$, $n = 1$, $\rho_{m0} = 1$, $m = 2$ and $\lambda = 5$}
\end{minipage}
\end{figure}
\begin{figure}[H]
\centering
\begin{minipage}[b]{0.4\textwidth}
    \includegraphics[width=\textwidth]{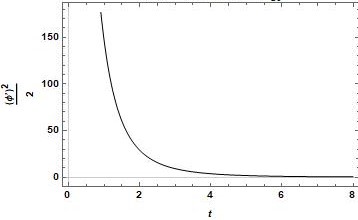}
    \caption{Plot of $\frac{1}{2}\dot{\phi}^2$ vs $t$ with $a_0 = 1$, $a_1 = 1$, $n = 1$, $\rho_{m0} = 1$, $m = 2$ and $\lambda = 5$}
\end{minipage}
\hfill
\begin{minipage}[b]{0.4\textwidth}
    \includegraphics[width=\textwidth]{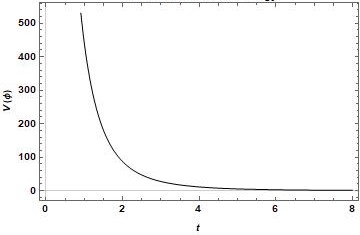}
    \caption{Plot of $V(\phi)$ vs $t$ with $a_0 = 1$, $a_1 = 1$, $n = 1$, $\rho_{m0} = 1$, $m = 2$ and $\lambda = 5$}
\end{minipage}
\end{figure}

\subsubsection{Energy conditions with reconstructed DBI-essence model}

Here we discuss the thermodynamic energy conditions. The
functional form of Scalar field energy density and pressure have
already been shown in the above section. Now the graphical
representations for energy conditions have been shown in Figs. 5
to 7. When time increases, Figs.5,6,7 show that the SEC is
violated while the NEC and DEC are satisfied.
\begin{figure}[H]
\centering
\begin{minipage}[b]{0.25\textwidth}
    \includegraphics[width=\textwidth]{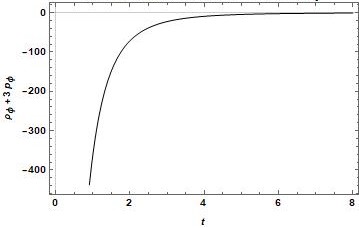}
    \caption{Plot of $\rho_{\phi}+3p_{\phi}$ vs $t$ with $a_0 = 1$, $a_1 = 1$, $n = 1$, $\rho_{m0} = 1$, $m = 2$ and $\lambda = 5$}
\end{minipage}
\hfill
\begin{minipage}[b]{0.25\textwidth}
    \includegraphics[width=\textwidth]{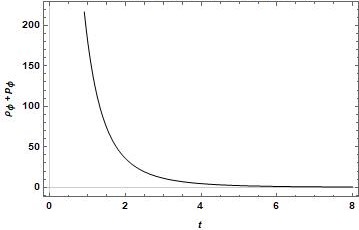}
    \caption{Plot of $\rho_{\phi}+p_{\phi}$ vs $t$ with $a_0 = 1$, $a_1 = 1$, $n = 1$, $\rho_{m0} = 1$, $m = 2$ and $\lambda = 5$}
\end{minipage}
\hfill
\begin{minipage}[b]{0.25\textwidth}
    \includegraphics[width=\textwidth]{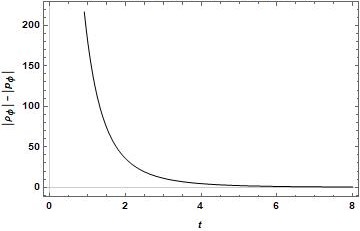}
    \caption{Plot of $\left |\rho_{\phi}\right |-\left |p_{\phi}\right |$ vs $t$ with $a_0 = 1$, $a_1 = 1$, $n = 1$, $\rho_{m0} = 1$, $m = 2$ and $\lambda = 5$}
\end{minipage}
\end{figure}

\subsubsection{Mass accretion formalism}

The basics of mass accretion have already been discussed in the
above section. Now we shall discuss the mass accretion graphically
for both Black holes and wormholes. The graphs have been shown as
follows in figs. 8 and 9. Fig.8 shows that the mass of the black
hole increases, while Fig.9 shows the decreasing nature of the
wormhole mass due to the accretion of the constructed DE.
\begin{figure}[H]
\centering
\begin{minipage}[b]{0.4\textwidth}
    \includegraphics[width=\textwidth]{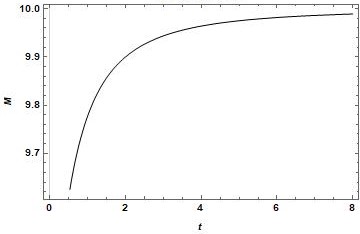}
    \caption{Plot of $M(t)$ vs $t$ for Blackholes with $a_0 = 1$, $a_1 = 1$, $n = 1$, $\rho_{m0} = 1$, $m = 2$ and $\lambda = 5$}
\end{minipage}
\hfill
\begin{minipage}[b]{0.4\textwidth}
    \includegraphics[width=\textwidth]{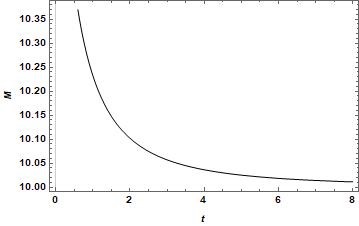}
    \caption{Plot of $M(t)$ vs $t$ for Wormholes with $a_0 = 1$, $a_1 = 1$, $n = 1$, $\rho_{m0} = 1$, $m = 2$ and $\lambda = 5$}
\end{minipage}
\end{figure}

\subsubsection{Reconstructed scalar field and potential}
Now we'll discuss the reconstructed scalar field and its potential
evolution w.r.t. the scalar field. This representation has been
done with using the power law scale factor. Representations have
been given in figs. 10 to 11.
\begin{figure}[H]
\centering
\begin{minipage}[b]{0.4\textwidth}
    \includegraphics[width=\textwidth]{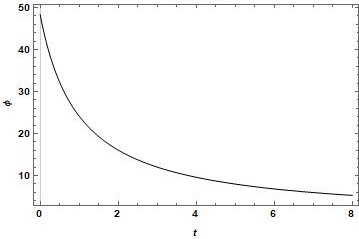}
    \caption{Plot of $\phi$ vs $t$ $a_0 = 1$, $a_1 = 1$, $n = 1$, $\rho_{m0} = 1$, $m = 2$ and $\lambda = 5$}
\end{minipage}
\hfill
\begin{minipage}[b]{0.4\textwidth}
    \includegraphics[width=\textwidth]{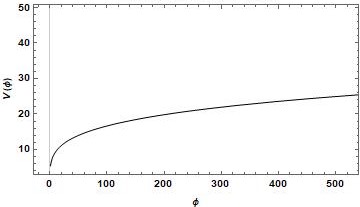}
    \caption{Plot of $V(\phi)$ vs $\phi$ with $a_0 = 1$, $a_1 = 1$, $n = 1$, $\rho_{m0} = 1$, $m = 2$ and $\lambda = 5$}
\end{minipage}
\end{figure}

\subsection{Scale factor $a(t)=g+a_0(a_1+nt)^m$}
The scale factor $a(t)=g+a_0(a_1+nt)^m$ is used in equations
(\ref{20}) and (\ref{21}) we get, the energy density $\rho_{\phi}$
and pressure $p_{\phi}$ which have been given in equations
(\ref{rh1}) and (\ref{p1}) in the \textbf{Appendix}.

This scale factor is also a special form of the first one, i.e.,
the power-law scale factor. This scale factor is capable of
resolving the past time singularity with $g=constant$. We are free
to choose the $g$ as a function of time, but for simplicity, in
the calculation, we have considered it as constant. For even power
$m$ the range of time should be $t\in (-\infty,+\infty)$.

So, those variables can be represented graphically as follows in
Figs. 12-15. From Fig. 12, we observe that the energy density
$\rho_{\phi}$ oscillates but takes finite value as time increases
from $t<0$ (i.e., before bounce) to $t>0$ (i.e., after bounce).
Also, before the bounce and after the bounce, Fig. 13 shows that
pressure $p_{\phi}$ is oscillating with a finite negative value.
We see that at $t=0$, both energy density and pressure are finite,
so these are regular in nature. From Figs. 14 and 15, we see that
the kinetic energy and potential energy both first increase upto
finite value and then decrease from a positive value to a low
positive value as time goes on ($t>0$).

\begin{figure}[H]
\centering
\begin{minipage}[b]{0.4\textwidth}
    \includegraphics[width=\textwidth]{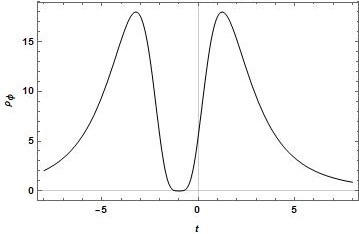}
    \caption{Plot of $\rho_{\phi}$ vs $t$ with $a_0 = 1$, $a_1 = 1$, $n = 1$, $\rho_{m0} = 1$, $m = 2$, $g = 5$ and $\lambda = 5$}
\end{minipage}
\hfill
\begin{minipage}[b]{0.4\textwidth}
    \includegraphics[width=\textwidth]{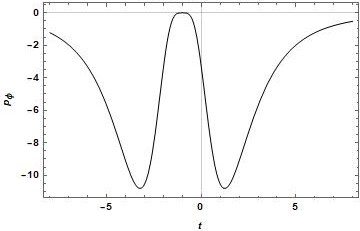}
    \caption{Plot of $p_{\phi}$ vs $t$ $a_0 = 1$, $a_1 = 1$, $n = 1$, $\rho_{m0} = 1$, $m = 2$, $g = 5$ and $\lambda = 5$}
\end{minipage}
\end{figure}
\begin{figure}[H]
\centering
\begin{minipage}[b]{0.4\textwidth}
    \includegraphics[width=\textwidth]{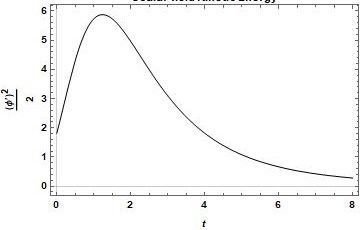}
    \caption{Plot of $\frac{1}{2}\dot{\phi}^2$ vs $t$ with $a_0 = 1$, $a_1 = 1$, $n = 1$, $\rho_{m0} = 1$, $m = 2$, $g = 5$ and $\lambda = 5$}
\end{minipage}
\hfill
\begin{minipage}[b]{0.4\textwidth}
    \includegraphics[width=\textwidth]{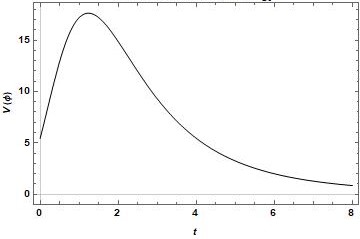}
    \caption{Plot of $V(\phi)$ vs $t$ with $a_0 = 1$, $a_1 = 1$, $n = 1$, $\rho_{m0} = 1$, $m = 2$, $g = 5$ and $\lambda = 5$}
\end{minipage}
\end{figure}

\subsubsection{Energy conditions with reconstructed DBI-essence model}

We can discuss the thermodynamics energy conditions. The
functional form of Scalar field energy density and pressure have
already been shown in the above section. Now the graphical
representations for energy conditions have been shown below in
Figs. 16 to 18. When time increases, Figs. 16-18 show that the SEC
is violated while the NEC and DEC are satisfied.
\begin{figure}[H]
\centering
\begin{minipage}[b]{0.25\textwidth}
    \includegraphics[width=\textwidth]{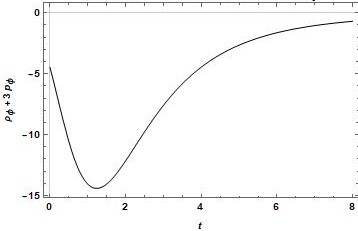}
    \caption{Plot of $\rho_{\phi}+3p_{\phi}$ vs $t$ with $a_0 = 1$, $a_1 = 1$, $n = 1$, $\rho_{m0} = 1$, $m = 2$, $g = 5$ and $\lambda = 5$}
\end{minipage}
\hfill
\begin{minipage}[b]{0.25\textwidth}
    \includegraphics[width=\textwidth]{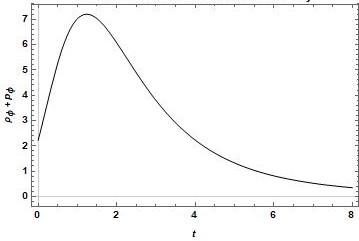}
    \caption{Plot of $\rho_{\phi}+p_{\phi}$ vs $t$ with $a_0 = 1$, $a_1 = 1$, $n = 1$, $\rho_{m0} = 1$, $m = 2$, $g = 5$ and $\lambda = 5$}
\end{minipage}
\hfill
\begin{minipage}[b]{0.25\textwidth}
    \includegraphics[width=\textwidth]{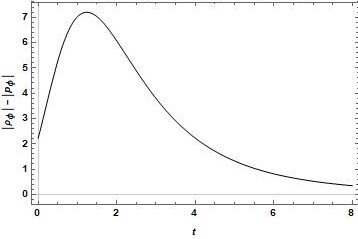}
    \caption{Plot of $\left |\rho_{\phi}\right |-\left |p_{\phi}\right |$ vs $t$ with $a_0 = 1$, $a_1 = 1$, $n = 1$, $\rho_{m0} = 1$, $m = 2$, $g = 5$ and $\lambda = 5$}
\end{minipage}
\end{figure}

\subsubsection{Mass accretion formalism}

The basics of mass accretion have already been discussed in the
above section. We will discuss the mass accretion graphically for
both Black holes and wormholes. The graphs have been shown as
follows in figs. 19 and 20. Fig.19 shows that the mass of the
black hole increases, while Fig.20 shows the decreasing nature of
the wormhole mass due to the accretion of the constructed DE.
\begin{figure}[H]
\centering
\begin{minipage}[b]{0.4\textwidth}
    \includegraphics[width=\textwidth]{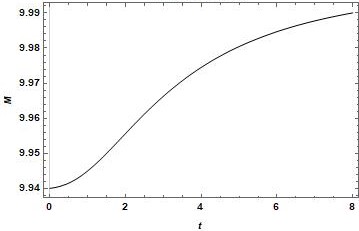}
    \caption{Plot of $M(t)$ vs $t$ for Blackholes with $a_0 = 1$, $a_1 = 1$, $n = 1$, $\rho_{m0} = 1$, $m = 2$, $g = 5$ and $\lambda = 5$}
\end{minipage}
\hfill
\begin{minipage}[b]{0.4\textwidth}
    \includegraphics[width=\textwidth]{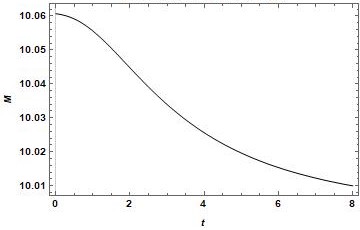}
    \caption{Plot of $M(t)$ vs $t$ for Wormholes with $a_0 = 1$, $a_1 = 1$, $n = 1$, $\rho_{m0} = 1$, $m = 2$, $g = 5$ and $\lambda = 5$}
\end{minipage}
\end{figure}

\subsection{Scale factor $a(t)=a_0\exp{\alpha t^2}$}

Using scale factor $a(t)=a_0\exp{\alpha t^2}$ \cite{24}in equations
(\ref{20}) and (\ref{21}) we get,
\begin{equation}
\rho_{\phi}=-\frac{\rho_{m0}\exp{(-3t^2 \alpha)}}{a^{3}_0}+1440t^4 \alpha^4 \lambda
\end{equation}

\begin{equation}
p_{\phi}=384t^2 \alpha^3 \lambda +288t^4 \alpha^4 \lambda + 48t^2 \alpha^2 (-8\alpha-24t^2\alpha^2)\lambda
\end{equation}

This scale factor has been considered to represent the inflation
with the cosmic bounce together. Here also the range should be
$t\in (-\infty,+\infty)$. So, those variables can be represented
graphically as follows in Figs. 21-24.

From Fig. 21, we observe that the energy density $\rho_{\phi}$
decreases and then increases from $t<0$ (i.e., before bounce) to
$t>0$ (i.e., after bounce). Also, before the bounce and after the
bounce, Fig. 22 shows that pressure $p_{\phi}$ increases and then
decreases but keeps negative value from $t<0$ (i.e., before
bounce) to $t>0$ (i.e., after bounce). We see that at $t=0$, both
energy density and pressure are very small, so these are regular
in nature. From Figs. 23 and 24, we see that the kinetic energy
and potential energy both increase but keep the positive value as
time goes on ($t>0$).

\begin{figure}[H]
\centering
\begin{minipage}[b]{0.4\textwidth}
    \includegraphics[width=\textwidth]{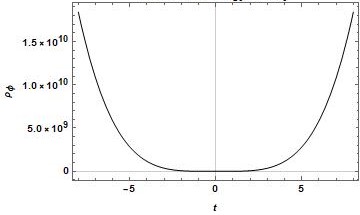}
    \caption{Plot of $\rho_{\phi}$ vs $t$ with $a_0 = 1$, $\rho_{m0} = 1$, $\alpha = 5$, and $\lambda = 5$}
\end{minipage}
\hfill
\begin{minipage}[b]{0.4\textwidth}
    \includegraphics[width=\textwidth]{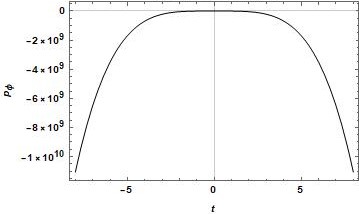}
    \caption{Plot of $p_{\phi}$ vs $t$ with $a_0 = 1$, $\rho_{m0} = 1$, $\alpha = 5$, and $\lambda = 5$}
\end{minipage}
\end{figure}
\begin{figure}[H]
\centering
\begin{minipage}[b]{0.4\textwidth}
    \includegraphics[width=\textwidth]{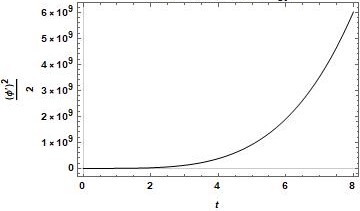}
    \caption{Plot of $\frac{1}{2}\dot{\phi}^2$ vs $t$ with $a_0 = 1$, $\rho_{m0} = 1$, $\alpha = 5$, and $\lambda = 5$}
\end{minipage}
\hfill
\begin{minipage}[b]{0.4\textwidth}
    \includegraphics[width=\textwidth]{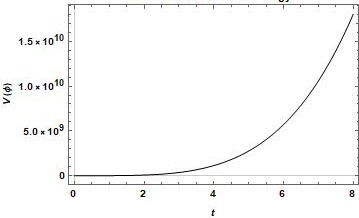}
    \caption{Plot of $V(\phi)$ vs $t$ with $a_0 = 1$, $\rho_{m0} = 1$, $\alpha = 5$, and $\lambda = 5$}
\end{minipage}
\end{figure}

\subsubsection{Energy conditions with reconstructed DBI-essence model}

Here we shall discuss the thermodynamics energy conditions. The
scalar field energy density and pressure found from the above
calculations by assuming the 2nd type of scale factor. The
functional form of Scalar field energy density and pressure have
already been shown in the above section. Now the graphical
representations for energy conditions have been shown in Figs. 25
to 27. When time increases, Figs. 25-27 show that the SEC is
violated while the NEC and DEC are satisfied.
\begin{figure}[H]
\centering
\begin{minipage}[b]{0.25\textwidth}
    \includegraphics[width=\textwidth]{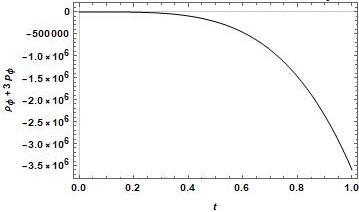}
    \caption{Plot of $\rho_{\phi}+3p_{\phi}$ vs $t$ with $a_0 = 1$, $\rho_{m0} = 1$, $\alpha = 5$, and $\lambda = 5$}
\end{minipage}
\hfill
\begin{minipage}[b]{0.25\textwidth}
    \includegraphics[width=\textwidth]{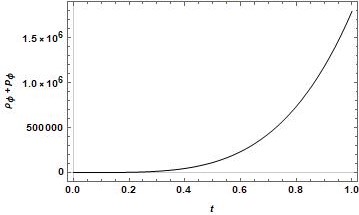}
    \caption{Plot of $\rho_{\phi}+p_{\phi}$ vs $t$ with $a_0 = 1$, $\rho_{m0} = 1$, $\alpha = 5$, and $\lambda = 5$}
\end{minipage}
\hfill
\begin{minipage}[b]{0.25\textwidth}
    \includegraphics[width=\textwidth]{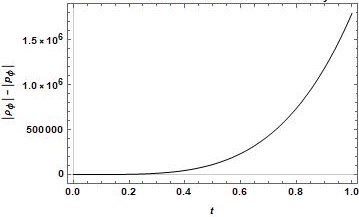}
    \caption{Plot of $\left |\rho_{\phi}\right |-\left |p_{\phi}\right |$ vs $t$ with $a_0 = 1$, $\rho_{m0} = 1$, $\alpha = 5$, and $\lambda = 5$}
\end{minipage}
\end{figure}

\subsubsection{Mass accretion formalism}

The basics of mass accretion have already been discussed in
section 6. Here in this section, we will discuss the mass
accretion graphically for both Black holes and wormholes. The
graphs have been shown as follows in Figs. 28 and 29. From these
figures, we see that the masses of the black hole and wormhole
decrease due to the accretion of the constructed DE.
\begin{figure}[H]
\centering
\begin{minipage}[b]{0.4\textwidth}
    \includegraphics[width=\textwidth]{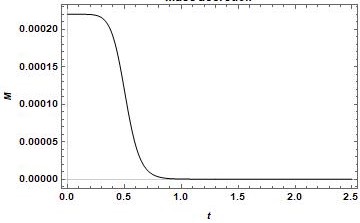}
    \caption{Plot of $M(t)$ vs $t$ for Blackholes with $a_0 = 1$, $\rho_{m0} = 1$, $\alpha = 5$, and $\lambda = 5$}
\end{minipage}
\hfill
\begin{minipage}[b]{0.4\textwidth}
    \includegraphics[width=\textwidth]{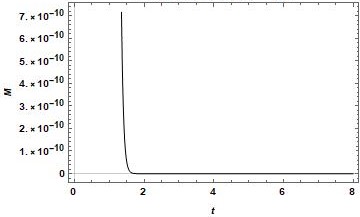}
    \caption{Plot of $M(t)$ vs $t$ for Wormholes with $a_0 = 1$, $\rho_{m0} = 1$, $\alpha = 5$, and $\lambda = 5$}
\end{minipage}
\end{figure}

\subsection{Scale factor $a(t)=a_0(\exp{\alpha t^2}+\exp{\alpha^2 t^4})$}

From scale factor $a(t)=a_0(\exp{\alpha t^2}+\exp{\alpha^2 t^4})$
\cite{24} in equations (\ref{20}) and (\ref{21}) we get, the
energy density $\rho_{\phi}$ and pressure $p_{\phi}$ which have
been given in equations (\ref{rh2}) and (\ref{p2}) in the
\textbf{Appendix}.

The scale factor provides the hybrid inflationary theory with
cosmic bounce. Here also the range should be $t\in
(-\infty,+\infty)$.

So, those variables can be represented graphically as follows in
Figs. 30-33. From Fig. 30, we observe that the energy density
$\rho_{\phi}$ decreases and then increases from $t<0$ (i.e.,
before bounce) to $t>0$ (i.e., after bounce). Also, before the
bounce and after the bounce, Fig. 31 shows that pressure
$p_{\phi}$ increases and then decreases but keeps negative value
from $t<0$ (i.e., before the bounce) to $t>0$ (i.e., after the
bounce). We see that at $t=0$, both energy density and pressure
are very small, so these are regular in nature. From Figs. 32 and
33, we see that the kinetic energy and potential energy both
increase but keep the positive value as time goes on ($t>0$).
\begin{figure}[H]
\centering
\begin{minipage}[b]{0.4\textwidth}
    \includegraphics[width=\textwidth]{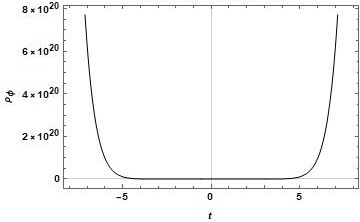}
    \caption{Plot of $\rho_{\phi}$ vs $t$ with $a_0 = 1$, $\rho_{m0} = 1$, $\alpha = 5$, and $\lambda = 5$}
\end{minipage}
\hfill
\begin{minipage}[b]{0.4\textwidth}
    \includegraphics[width=\textwidth]{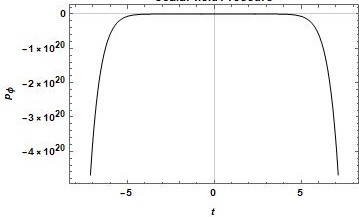}
    \caption{Plot of $p_{\phi}$ vs $t$ with $a_0 = 1$, $\rho_{m0} = 1$, $\alpha = 5$, and $\lambda = 5$}
\end{minipage}
\end{figure}
\begin{figure}[H]
\centering
\begin{minipage}[b]{0.4\textwidth}
    \includegraphics[width=\textwidth]{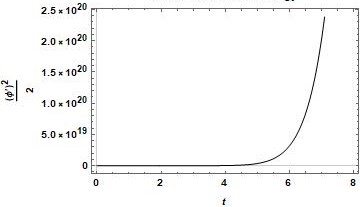}
    \caption{Plot of $\frac{1}{2}\dot{\phi}^2$ vs $t$ with $a_0 = 1$, $\rho_{m0} = 1$, $\alpha = 5$, and $\lambda = 5$}
\end{minipage}
\hfill
\begin{minipage}[b]{0.4\textwidth}
    \includegraphics[width=\textwidth]{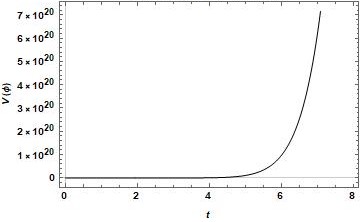}
    \caption{Plot of $V(\phi)$ vs $t$ with $a_0 = 1$, $\rho_{m0} = 1$, $\alpha = 5$, and $\lambda = 5$}
\end{minipage}
\end{figure}

\subsubsection{Energy conditions with reconstructed DBI-essence model}

The functional form of Scalar field energy density and pressure
have already been shown in above section. Now the graphical
representations for energy conditions have been shown in Figs. 34
to 36. When time increases, Figs. 34-36 show that the SEC is
violated while the NEC and DEC are satisfied.
\begin{figure}[H]
\centering
\begin{minipage}[b]{0.25\textwidth}
    \includegraphics[width=\textwidth]{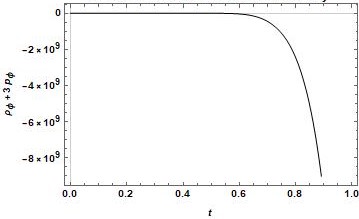}
    \caption{Plot of $\rho_{\phi}+3p_{\phi}$ vs $t$ with $a_0 = 1$, $\rho_{m0} = 1$, $\alpha = 5$, and $\lambda = 5$}
\end{minipage}
\hfill
\begin{minipage}[b]{0.25\textwidth}
    \includegraphics[width=\textwidth]{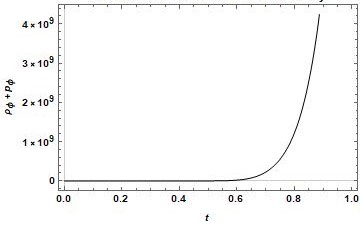}
    \caption{Plot of $\rho_{\phi}+p_{\phi}$ vs $t$ with $a_0 = 1$, $\rho_{m0} = 1$, $\alpha = 5$, and $\lambda = 5$}
\end{minipage}
\hfill
\begin{minipage}[b]{0.25\textwidth}
    \includegraphics[width=\textwidth]{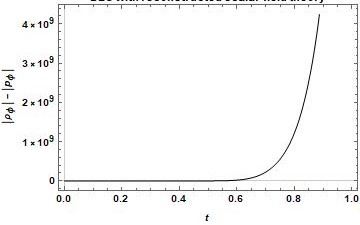}
    \caption{Plot of $\left |\rho_{\phi}\right |-\left |p_{\phi}\right |$ vs $t$ with $a_0 = 1$, $\rho_{m0} = 1$, $\alpha = 5$, and $\lambda = 5$}
\end{minipage}
\end{figure}

\subsubsection{Mass accretion formalism}

Now we will discuss the mass accretion graphically for both Black
holes and wormholes, as discussed in section 6. The graphs have
been shown as follows in Figs. 37 and 38. From these figures, we
see that the mass of the black hole decreases while wormhole mass
first slightly increases upto $t\approx 0.3$ and then decreases
due to accretion of the constructed DE.
\begin{figure}[H]
\centering
\begin{minipage}[b]{0.4\textwidth}
    \includegraphics[width=\textwidth]{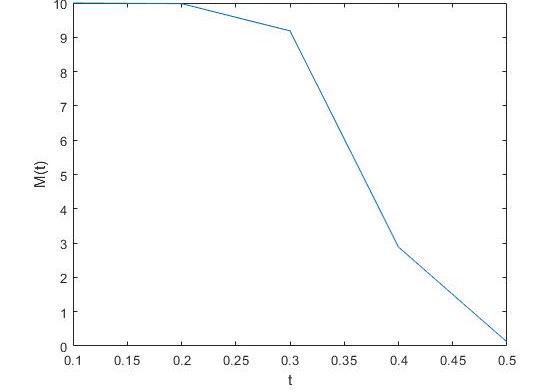}
    \caption{Plot of $M(t)$ vs $t$ for Black holes with $a_0 = 1$, $\rho_{m0} = 1$, $\alpha = 5$, and $\lambda = 5$}
\end{minipage}
\hfill
\begin{minipage}[b]{0.4\textwidth}
    \includegraphics[width=\textwidth]{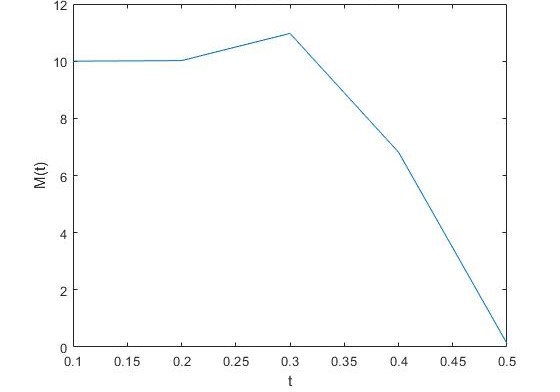}
    \caption{Plot of $M(t)$ vs $t$ for Wormholes with $a_0 = 1$, $\rho_{m0} = 1$, $\alpha = 5$, and $\lambda = 5$}
\end{minipage}
\end{figure}

\section{Physical analysis}
In this work, we have considered four different types of scale
factor solutions of Friedmann equations. We have analyzed energy
density, pressure, scalar field kinetic energy, potential energy,
thermodynamic energy conditions with density-pressure approach,
and mass accretion for black holes. and wormholes with these four
types of scale factors.\par In figures 1, 2, 12, 13, 21, 22, 30
and 31, we have observed that all the four scale factors give
positive energy density and negative pressure. These results
support the observations of expanding universe. The scale factors
used in this work show a concave up monotonic nature from which we
can conclude that our models also support the accelerated
expansion of the universe.\par For the power-law type of scale
factors (1st and 2nd), we have observed that the kinetic energy
and potential energy both decrease with time. Particularly for the
second scale factor, both energies increase at an early time and
decrease at the late time (Figs. 14 and 15). But for the first
type scale factor, both energies remain totally monotonic during
both early and late time periods (Figs. 3 and 4). For the last two
scale factors i.e., the inflationary scale factors (bouncing
exponential and bouncing hybrid-exponential scale factor), we have
found increasing nature of scalar field kinetic energies and
potential energies (Figs. 23, 24, 32 and 33).\par In the analysis
of thermodynamic energy conditions (Figs. 5-7, 16-18, 25-27 and
34-36), we have observed that for the first type of scale factor,
the SEC is violated, but other energy conditions are satisfied
(Fig. 5 to 7). For the second type of scale factor, in the early
time phase, the results for energy conditions are similar to that
of the first type of scale factor. But at the late time phase, all
the energy conditions are satisfied with the second type of scale
factor (Fig. 16 to 18). In the last two inflationary scale factors
(bouncing exponential and bouncing hybrid-exponential scale
factor), we have observed that the strong energy conditions are
violated, but the other energy conditions are satisfied (Figs. 25
to 27 and 34 to 36).\par For the first two types of power-law
scale factors, the black hole mass increases, and the rate of mass
accretion also increases. On the other hand, the mass of the
wormhole decreases with time. For the last two inflation-type
bouncing scale factors, the masses of both black holes and
wormholes decrease with time. From this, we can conclude that
during late time accelerated expanding universe, the effect of
gravity of black holes should be increased with time, and that of
wormholes would vanish with time. The first two types of scale
factors provide solutions for late time expanding model. That is
why it is difficult to observe wormholes experimentally. On the
other hand, during inflations, the black holes and wormholes both
should vanish, and hence it is difficult to find primordial black
holes as well as wormholes experimentally at present time universe
(Figs. 8, 9, 19, 20, 28, 29, 37 and 38).\par For the first type of
scale factor, we have represented the time evolution of scalar
field and evolution of scalar field potential with respect to a
scalar field. From the potential, it is observed that after a
bounce, the universe should run through a slow-roll inflationary
stage. The scalar field decreases continuously, and this can
provide late-time warm inflation in the universe (Figs. 10 and
11). \par Particularly, in the application of scale factors, we
have chosen the constants and parameters (power parameter $m$,
multiplicity parameter $n$ and order parameter $g$ for the first
two power-law type solutions and multiplicity parameter $\alpha$
for the last two type bouncing inflationary solutions) such that
we can find a non-singular point at $t\rightarrow 0$. For the last
two type solutions, the universe represent symmetric nature (with
respect to scale factor) at $t = 0$. For the first two scale
factors universe shows asymmetric nature in scale factor
evolution.

\section{Concluding remarks}
We have discussed the reconstructed DBI-essence model with the
introduction of four types of scale factors. Each scale factor
represents some different cosmic phases. We have discussed the
reconstructed scalar field energy density, pressure, kinetic
energy, and potential energy for all those scale factors. Most of
the time, we found negative pressure that provides proof of
accelerated expansion of the universe.\par We have discussed the
energy conditions for all those scale factors. This paper compares
the energy conditions and their variations for different cosmic
phases. \par The mass accretion also provides the idea of black
holes and wormholes mass change in different cosmic phases. The
first scale factor represents the late time expanding phase. The
second scale factor represents the bouncing cosmic phase. The last
two provide the existence of cosmic inflation by introducing the
bounce. \par Overall, we have tried to establish the DBI dark
energy cosmology under reconstruction formalism. Although the
DBI-essence is one kind of most generalized scalar field theory,
our reconstructed model generalized it further which helps to
represent complete evolutionary scheme of the universe.\\\\

\textbf{Appendix}\\

Using the scale factor given in section (7.2), we get the energy
density and pressure as
\begin{equation}\label{rh1}
    \rho_{\phi}=-\frac{\rho_{m0}}{(g+a_0(a_1+nt)^m)^3}+\frac{90m^4 n^4 a^{4}_0 \lambda (a_1+nt)^{-4+4m}}{(g+a_0(a_1+nt)^m)^4}
\end{equation}
and
\begin{equation}\label{p1}
\begin{split}
    p_{\phi}=(\frac{18m^4n^4a^{4}_0 (a_1+nt)^{-4+4m}\lambda}{(g+a_0(a_1+nt)^m)^4})\\+(\frac{4mna_{0}(a_1+nt)^{-1+m}(-\frac{12m^3n^3a^{3}_0 (a_1+nt)^{-3+3m}}{(g+a_0(a_1+nt)^m)^3}+\frac{6m^2n^3(-2+2m)a^{2}_0 (a_1+nt)^{-3+2m}}{(g+a_0(a_1+nt)^m)^2})\lambda}{g+a_0(a_1+nt)^m})\\+(\frac{12m^2n^2a^{2}_0(a_1+nt)^{-2+2m}(-\frac{6m^2n^2a^{2}_0 (a_1+nt)^{-2+2m}}{(g+a_0(a_1+nt)^m)^2}-4(-\frac{m^2n^2a^{2}_0 (a_1+nt)^{-2+2m}}{(g+a_0(a_1+nt)^m)^2}+-\frac{(-1+m)mn^2a_0 (a_1+nt)^{-2+m}}{(g+a_0(a_1+nt)^m)}))}{(g+a_0(a_1+nt)^m)^2})
\end{split}
\end{equation}

Using the scale factor given in section (7.4), we get the energy
density and pressure as
\begin{equation}\label{rh2}
\rho_{\phi}=-\frac{\rho_{m0}}{(\exp{(t^2 \alpha)}+\exp{(t^4
\alpha^2)})^3 a^{3}_0}+\frac{90\lambda (2\exp{(t^2
\alpha)}t\alpha+4\exp{(t^4
\alpha^2)}t^3\alpha^2)^4\lambda}{(\exp{(t^2 \alpha)}+\exp{(t^4
\alpha^2)})^4}
\end{equation}
and
\begin{equation}\label{p2}
\begin{split}
    p_{\phi}=(\frac{18\lambda(2e^{\alpha t^2}t\alpha+4e^{t^4\alpha^2}t^3\alpha^2)^4}{(e^{t^2\alpha}+e^{t^4\alpha^2})^4})\\+(\frac{4(2e^{\alpha t^2}t\alpha+4e^{t^4\alpha^2}t^3\alpha^2)(-\frac{12\lambda(2e^{\alpha t^2}t\alpha+4e^{t^4\alpha^2}t^3\alpha^2)^3}{(e^{t^2\alpha}+e^{t^4\alpha^2})^3}+\frac{12(2e^{\alpha t^2}t\alpha+4e^{t^4\alpha^2}t^3\alpha^2)(2e^{\alpha t^2}\alpha+4e^{\alpha t^2}t^2\alpha^2+12e^{t^4 \alpha^2}t^2\alpha^2+16e^{t^4 \alpha^2}t^6\alpha^4)}{(e^{t^2\alpha}+e^{t^4\alpha^2})^2})\lambda}{e^{t^2\alpha}+e^{t^4\alpha^2}})\\+(\frac{A(B+C)}{D})
\end{split}
\end{equation}

where $A$, $B$, $C$ and $D$ are as follows.
\begin{center}
    $A=12\lambda(2e^{\alpha t^2}t\alpha+4e^{t^4\alpha^2}t^3\alpha^2)^2$
\end{center}
\begin{center}
    $B=-\frac{6\lambda(2e^{\alpha t^2}t\alpha+4e^{t^4\alpha^2}t^3\alpha^2)^2}{(e^{t^2\alpha}+e^{t^4\alpha^2})^2}$
\end{center}
\begin{center}
    $C=-4(-\frac{(2e^{\alpha t^2}t\alpha+4e^{t^4\alpha^2}t^3\alpha^2)^2}{(e^{t^2\alpha}+e^{t^4\alpha^2})^2}+\frac{(2e^{\alpha t^2}\alpha+4e^{\alpha t^2}t^2\alpha^2+12e^{t^4 \alpha^2}t^2\alpha^2+16e^{t^4 \alpha^2}t^6\alpha^4)}{e^{t^2\alpha}+e^{t^4\alpha^2}})$
\end{center}
\begin{center}
    $D=(e^{t^2\alpha}+e^{t^4\alpha^2})^2$
\end{center}

\end{document}